\documentclass[prl,reprint,twocolumn,showpacs,superscriptaddress]{revtex4}
\usepackage{graphicx}
\usepackage{amsmath}
\usepackage{amssymb}
\usepackage{bm}
\renewcommand{\vec}[1]{\bm{#1}}

\begin{document}

\title{The Coefficient of Restitution as a Fluctuating Quantity}

\author{Marina Montaine}

\author{Michael Heckel}
\affiliation{Institute for Multiscale Simulation, Friedrich-Alexander Universit\"at Erlangen-N\"urnberg, Erlangen, Germany}

\author{Christof Kruelle}
\affiliation{Hochschule Karlsruhe, Fakult\"at f\"ur Maschinenbau und Mechatronik, Karlsruhe, Germany}

\author{Thomas Schwager}
\affiliation{Bruker Nano GmbH, Berlin, Germany}

\author{Thorsten P\"oschel}
\affiliation{Institute for Multiscale Simulation, Friedrich-Alexander Universit\"at Erlangen-N\"urnberg, Erlangen, Germany}

\date{\today}

\begin{abstract}
The coefficient of restitution of a spherical particle in contact with a flat plate is investigated as a function of the impact velocity. As an experimental observation we notice non-trivial (non-Gaussian) fluctuations of the measured values. For a fixed impact velocity, the probability density of the coefficient of restitution, $p(\varepsilon)$, is formed by two exponential functions (one increasing, one decreasing) of different slope. This behavior may be explained by a certain roughness of the particle which leads to energy transfer between the linear and rotational degrees of freedom. 
\end{abstract}



\maketitle

\paragraph{Introduction.} The dissipative collision of a solid particle with a hard plane may be described by the coefficient of normal restitution
\begin{equation}
\varepsilon = -\dfrac{\vec{v}^{\,\prime}\cdot\vec{n}}{\vec{v}\cdot\vec{n}}\,,
\end{equation}
relating the normal components of the relative velocity before and after a collision at the point of contact. The unit vector $\vec{n}$ indicates the relative particle positions at the instant of the collision. Obviously, $\varepsilon=1$ stands for elastic collisions whereas $\varepsilon=0$ indicates the complete dissipation of the energy of the relative motion. There are several techniques for measuring the coefficient of restitution, including high-speed video analysis, e.g., \cite{Labous:1997} and sophisticated techniques, where the particle is attached to a compound pendulum with the axis of rotation very close to the center of mass \cite{Bridges:1984,Hatzes:1988} which makes this method particularly suitable for the measurement of the coefficient of restitution for very small impact velocities up to mm/sec and below. In the presence of gravity, the coefficient of restitution can be measured by determining the time lag between consecutive impacts of a particle bouncing on a hard plane, using a piezoelectric force sensor, e.g., \cite{Koller:1987,Falcon:1998} or an accelerometer mounted to the plate which detects elastic waves excited by the impact, e.g., \cite{King:2005}. When both particle and plate are metallic, the time of the impacts can be determined by applying a DC voltage between ball and plate and determining the instant when the circuit closes \cite{King:2005}. In {\em many} papers, the time lag and, thus, $\varepsilon$ is determined by recording the sound emitted from a spherical particle bouncing on an underlying flat plane, e.g., \cite{Bernstein:1977,Stensgaard:2001} and many others. From the times $t_i$ of impacts one can compute the coefficient of restitution via
\begin{equation}
 \label{eq:epsmeas}
 \varepsilon(v_i)=\frac{v_i^\prime}{v_i} = \frac{t_{i+1}-t_{i}}{t_{i}-t_{i-1}}\,,~~~v_i=\frac{g}{2}(t_i-t_{i-1})\,,
\end{equation}
where $v_i$ and $v_i^\prime$ are the normal pre- and postcollisional velocities of the impact taking place at time $t_i$ and $g$ is the acceleration due to gravity.

Although it is frequently assumed that the coefficient of restitution is a material constant (e.g., this assumption is common in many textbooks and widely used in simulations), numerous experimental studies have revealed that it depends on many parameters: impact velocity, material characteristics of the impacted bodies, particle size, shape and roughness, and adhesion properties. Here we restrict ourselves to the investigation of $\varepsilon$ of a single dry steel ball bouncing on a hard plane, such that the coefficient of restitution is only a function of the impact velocity $\varepsilon=\varepsilon(v)$.

In several experimental investigations, using either photographic techniques \cite{Sorace:2009} or an acoustic emission analysis \cite{Falcon:1998}, an extraordinary high fluctuation level was noticed whose origin remained obscure and cannot be attributed to the imperfections of the experimental setup. By means of large-scale experiments as well as micromechanical modeling we make an attempt to characterize the fluctuations of the coefficient of restitution and to explain the mechanism leading to these fluctuations.

\paragraph{Experiments.} Our experimental setup consists of a robot to move a small vertical tube to a desired position $\{x,y,z\}$. In the beginning of each experimental trial a stainless steel bearing ball is suspended at the end of the tube by means of a vacuum pump. Switching off the pump the sphere is released to bounce repeatedly off the ground where the sound is recorded by a piezoelectric sensor. When the ball eventually comes to rest, it is pushed to a defined position by a fan where it is picked up by the robot who moves the ball again to the start position for the next trial. In each cycle the initial $\{x,y\}$-position is chosen randomly within the central region of the ground plate such that edge-effects \cite{Sondergaard:1990} are not noticeable. The dropping height is chosen randomly from $z \in [9, 10]$\,cm, corresponding to impact velocities $v\in[1.33, 1.4]$\,m/s which allows for the observation of 90-100 bounces of the ball. A massive hard glass plate of size 30$\times$20$\times$1.9\,cm$^3$ can either serve directly as a ground plate for the experiment or serve as a carrier for other ground plates. For our experiments we used an extremely hard SiC disk and a Vitreloy disk. The measuring process is fully automatized, this allows us to perform a large-scale experiment collecting data from thousands of bouncing ball trials.

\begin{figure}[htbp]
 \centerline{\includegraphics[width=0.9\columnwidth,clip]{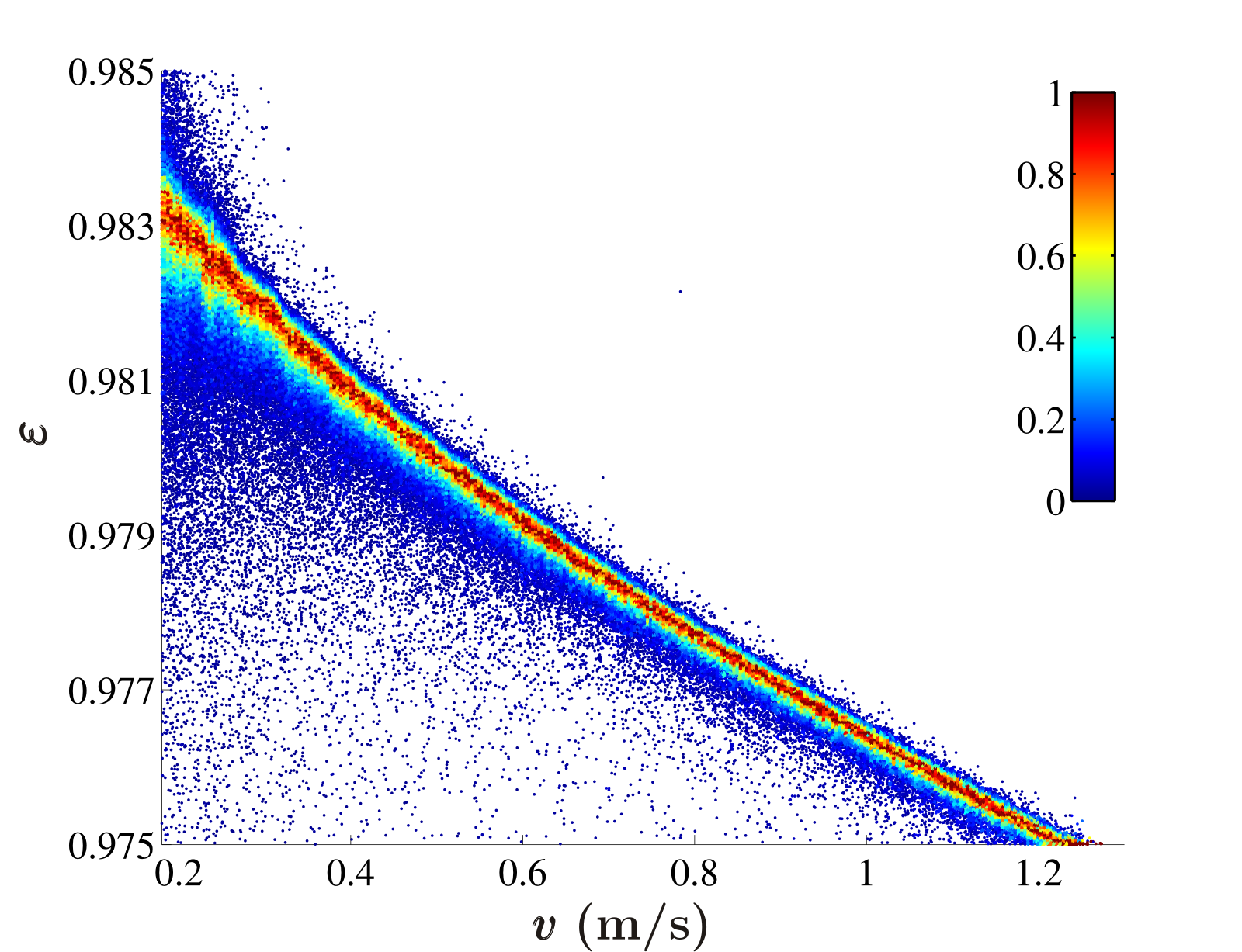}}
 \centerline{\includegraphics[width=0.9\columnwidth,clip]{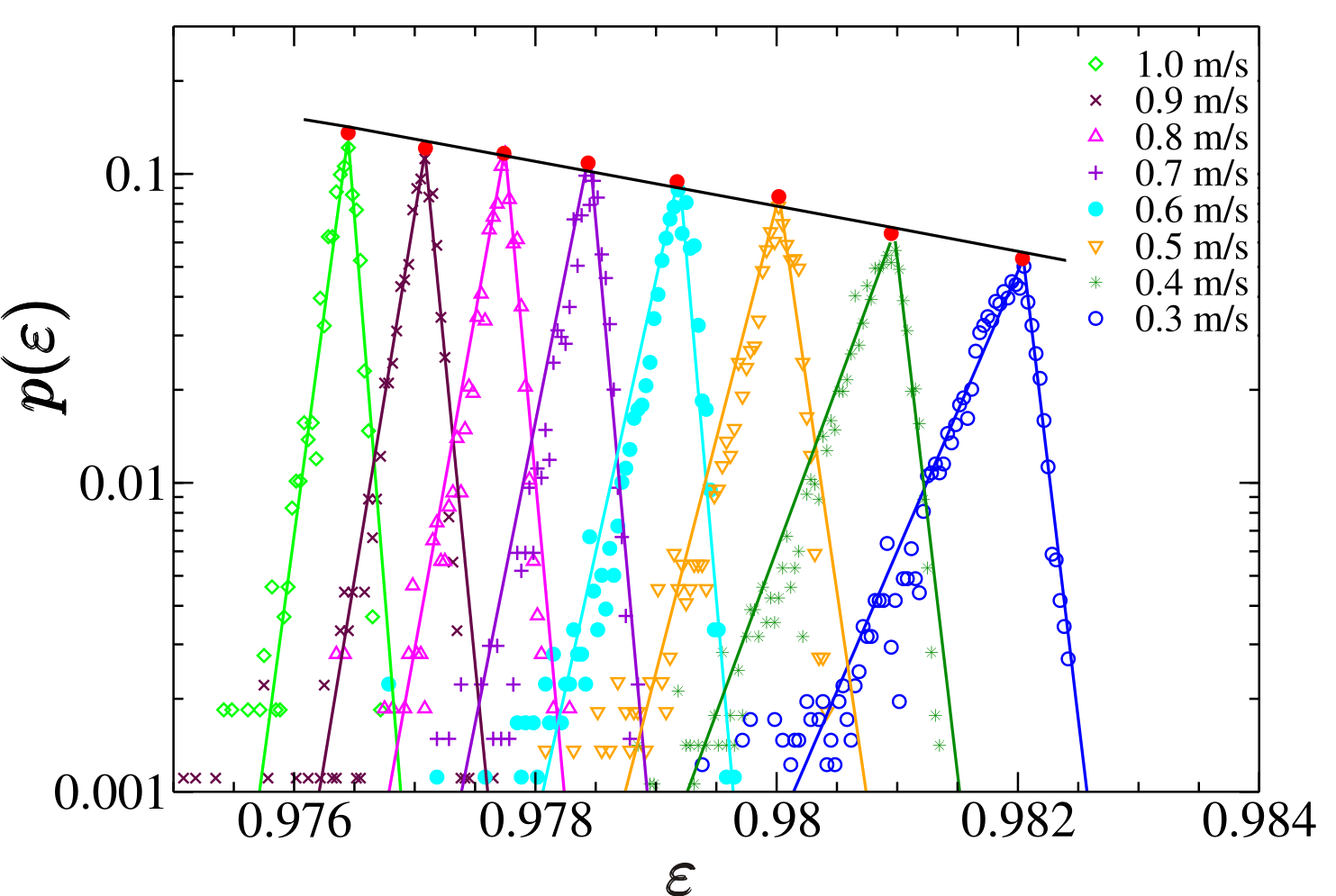}}
\caption{(color online). Experimental results. Top: Coefficient of restitution $\varepsilon$ plotted against the normal impact velocity $v$. The data are colored according to the normalized frequency of occurrences. Bottom: Histograms of $\varepsilon$ for impact velocities from small intervals, centered around $v=0.3\:\ldotp\ldotp1.0$ m/s. The lines are exponential fits.}
 \label{fig:Exp}
\end{figure}

From the analysis of the sound-sensor signal we obtain the impact times $t_i$ and, thus, via Eq. \eqref{eq:epsmeas} the normal impact velocities $v_i$ and the coefficients of restitution $\varepsilon(v_i)$.
Figure \ref{fig:Exp} displays the abundance of data, $\varepsilon(v)$, for a stainless steel ball (radius $R=3.0$ mm, Young modulus $Y=200$ GPa, Poisson ratio $\nu=0.30$, density $\rho_s=7.90 $ g/cm$^3$). Besides the expected decay of $\varepsilon$ with increasing impact velocity, e.g., \cite{Brilliantov:1996,SchwagerPoeschel:2008}, we observe a large scatter of the data. This scatter becomes apparent only when analyzing a large number of measurements (here 220,000) and was, therefore, not noticed in earlier, similar experiments \cite{Koller:1987,Falcon:1998}.
The scatter of $\varepsilon$ is asymmetric, that is, the deviation of the data with $\varepsilon$ lower than the mean is noticeably larger. Figure \ref{fig:Exp} (bottom) shows the normalized frequency $p(\varepsilon)$ of measurements of a certain value of $\varepsilon$ for several small intervals of the impact velocity $v$. Thus, if we consider $\varepsilon$ as a fluctuating quantity,  $p(\varepsilon)$ may be considered as its probability density function. This function reveals strongly non-Gaussian behavior, but the distribution is shaped by a combination of two exponential functions (one increasing, one decreasing) of different slope. We believe that uncommon statistical properties are not due to imperfections of the experimental setup but are a consequence of microscopic asperities at the sphere's surface. This hypothesis is checked by means of a numerical simulation.

\paragraph{Numerical simulations.} The procedure of our simulation is analogue to the experiment; the rigid ball is dropped from a certain height $h$ and repeatedly collides with a smooth, hard plate. The air drag is neglected and the ball is subjected only to gravity.

\begin{figure}[htbp!]
\centering
\includegraphics[width=0.45\columnwidth]{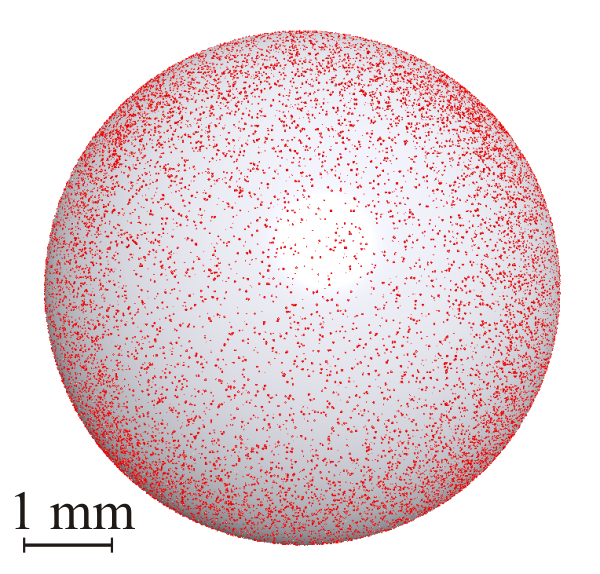}
\includegraphics[width=0.45\columnwidth]{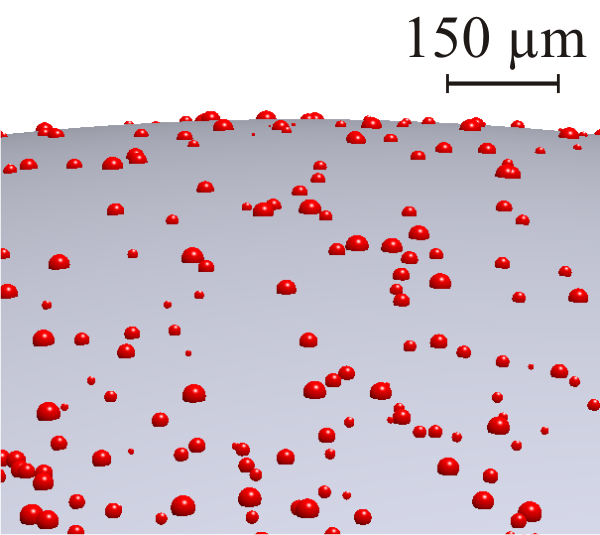}
\caption{Sketch of the particle model and closeup of its surface.}
\label{fig:ball}
\end{figure}

The ball is modeled as a composite multi-sphere particle (Fig. \ref{fig:ball}).
The large central sphere of radius $R$ is randomly covered by many ($N\sim 10^6$) tiny spheres of different microscopic size, representing asperities. The center of each small sphere $i$ of radius $R_i$ is located at the surface of the central sphere. Since $R_i/R\sim 10^{-3}$ (mass ratio $m_i/m\sim 10^{-9}$) we can safely neglect the contribution of the small particles for the computation of the moment of inertia of the ball. Thus, the ball is characterized by its mass $m$, $R$, center of mass velocity $\vec{v}_G$, angular velocity $\vec{\omega}$, the set of the radii, $R_i$, of the asperities and the set of their position vectors, $\vec{r}_i$, pointing from the center of the large sphere to the asperities. When bouncing, the ball may come in contact with the plane either through the central sphere or through one of the asperities. In both cases the collision is assumed instantaneous and inelastic, characterized by coefficients of restitution.

\begin{figure}[htbp]
 \centerline{\includegraphics[width=0.8\columnwidth,clip]{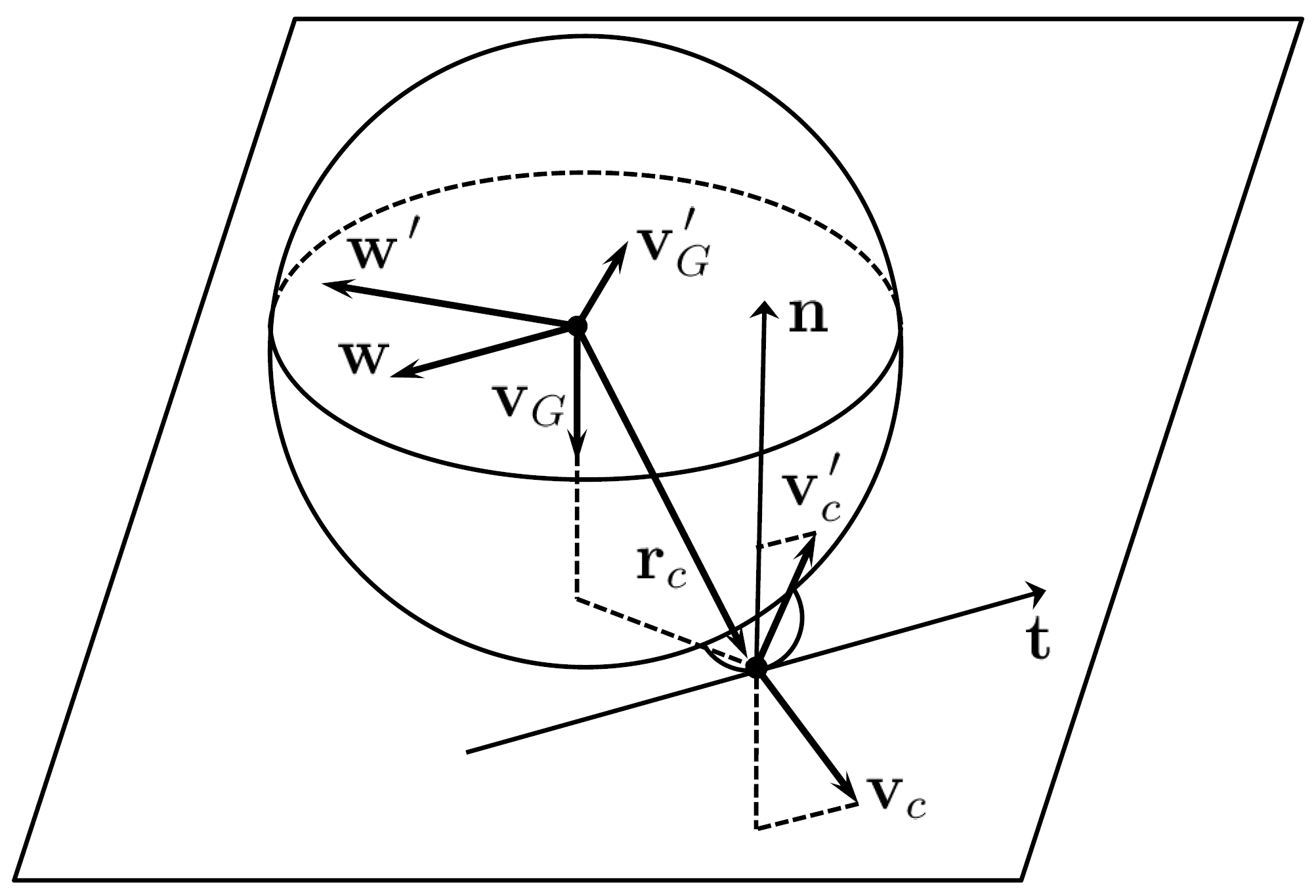}}
\caption{Sketch of a particle collision. For simplicity only the impacting asperity (of exaggerated size) is drawn.}
 \label{fig:sketch}
\end{figure}

In the intervals between collisions the ball follows a ballistic trajectory
\begin{equation}
 \label{eq:rG}
 \vec{r}_G(t)=\vec{r}_G(t_0) + \vec{v}_G(t_0)\,(t-t_0) + \dfrac{(t-t_0)^2}{2}\vec{g}, \quad t\geq t_0  
\end{equation}
where $t_0$ is the time of the preceding collision and $\vec{g}$ is gravity while the angular velocity $\vec{\omega}\equiv \omega\vec{e}_\omega$ remains constant.
The evolution of a vector $\vec{p}$ fixed to the particle is then given by
\begin{equation}
  \begin{split}
    \vec{p}_\text{rot} & =\vec{p} \cos\omega t + \vec{e}_\omega(\vec{e}_\omega \cdot \vec{p}\,) (1 - \cos\omega t) +(\vec{e}_\omega \times \vec{p}\,)\sin\omega t\\
&=\hat{A}(t)\vec{p},
\end{split}
\label{eq:r_rot}
\end{equation}
which defines the rotation matrix $\hat{A}$ \cite{Goldstein:2001}.
Therefore, the particle contacts the ground when the center of the first asperity $j$ reaches the height $R_j$, that is, the time of contact $t_\text{c}$ follows from the condition
\begin{equation}
  \label{eq:mincondition}
  \min_{j=1,N}\left[\vec{r}_G(t_c) + \hat{A}(t_c)\vec{r}_j \right]_z = R_j\,,
\end{equation}
where $\left[...\right]_z$ means the vertical component of the argument. The vectorial impact velocity at the point of contact $\vec{r}_c$ is then
 \begin{equation}
   \label{eq:vc}
   \vec{v}_c = \left(\vec{v}_c\cdot \vec{n}\right)\vec{n} + \left(\vec{v}_c\cdot \vec{t}\,\,\right)\vec{t} = \vec{v}_G+\vec{\omega}\times \vec{r}_c \,,
 \end{equation}
where $\vec{n}$ and $\vec{t}$ are the unit vectors in normal and tangential directions,
see Fig. \ref{fig:sketch}.

The post-collisional velocity at the contact point is given by
\begin{equation}
  \label{eq:epsdef}
  \vec{v}^{\,\prime}_c\cdot\vec{n}  =  -\varepsilon \left(\vec{v}_c\cdot\vec{n}\right)\,,~~~~~
  \vec{v}^{\,\prime}_c\cdot\vec{t} =  \beta \left(\vec{v}_c\cdot\vec{t}\,\,\right)\,,
\end{equation}
with the coefficients of normal and tangential restitution $\varepsilon$ and $\beta$. Finally, we compute the post-collisional velocity $\vec{v}_G^{\,\prime}$ and angular velocity $\vec{\omega}^\prime$
\begin{equation}
 \label{eq:P}
 \begin{split}
\vec{v}_G^{\,\prime} -\vec{v}_G & = \Delta \vec{v}_G = \frac{1}{m}\vec{P}\\
\vec{\omega}^\prime - \vec{\omega} & = \Delta\vec{\omega} = \frac{1}{\hat{J}}\vec{r}_c\times \vec{P}\,,
\end{split}
\end{equation}
where the transferred momentum $\vec{P}$ is obtained from Eqs. \eqref{eq:vc} and \eqref{eq:epsdef}:
\begin{equation}
 \label{eq:deltaVc}
 \vec{v}_c^{\,\prime}-\vec{v}_c = \Delta\vec{v}_G+\Delta \vec{\omega}\times \vec{r}_c=\frac{1}{m}\vec{P}+\frac{1}{\hat{J}}\left(\vec{r}_c\times \vec{P}\right)\times \vec{r}_c
\end{equation}
with the mass $m$ and the moment of inertia $\hat{J}$ of the particle.

Using Eqs. \eqref{eq:mincondition} and \eqref{eq:P} we can compute the dynamics of the bouncing particle and, in particular, the times and velocities of the impacts, while the coefficients of restitution, $\varepsilon$ and $\beta$ are given. For $\varepsilon$ as a function of the normal impact velocity ${v}_c$ we use the expression for viscoelastic spheres \cite{SchwagerPoeschel:2008}
\begin{equation}
 \label{eq:epsvisco}
 \varepsilon({v}_c)=1+\sum_{i=1}^\infty C_iA_i({v}_c)^{i/10},
\end{equation}
with $C_i$ being known constants ($C_1=C_3=0$; $C_2=-1.153$; $C_4=0.798$; $C_5=0.267$\dots, see \cite{SchwagerPoeschel:2008} for details) and $A_i$ being material constants. We determined the first three non-trivial constants, $A_2\approx 0.0467$; $A_4\approx 0.1339$; $A_5\approx -0.2876$, by fitting Eq. \eqref{eq:epsvisco} to the experimental data shown in Fig. \ref{fig:Exp}.

The coefficient of tangential restitution $\beta$ depends on both bulk material properties and surface properties. Therefore, $\beta$ cannot be analytically derived from material properties, except for the limiting case of pure Coulomb friction \cite{SchwagerBecker:2008}. Here we use $\beta=1$.

\begin{figure}[htbp]
 \centerline{\includegraphics[width=0.9\columnwidth,clip]{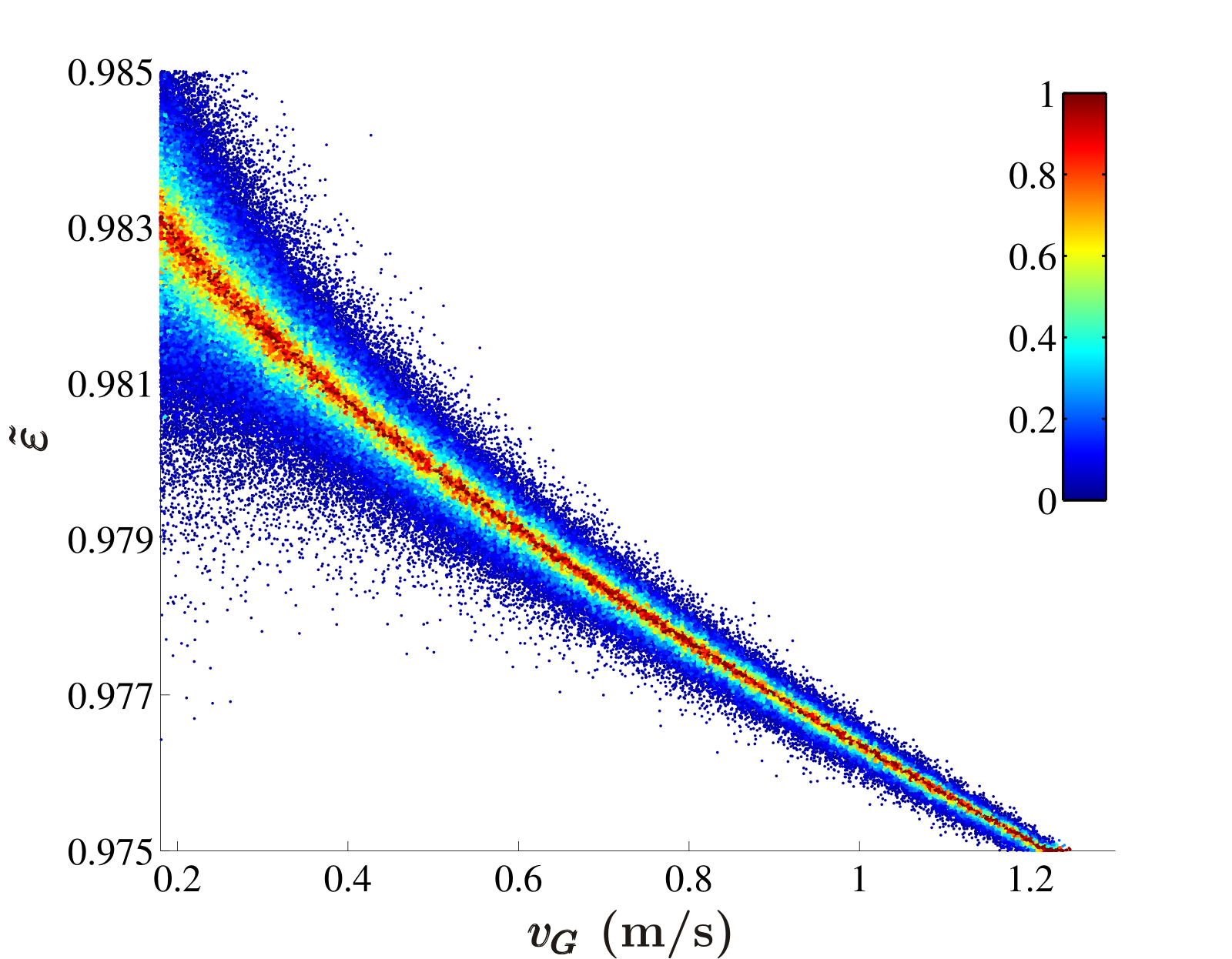}}
 \centerline{\includegraphics[width=0.9\columnwidth,clip]{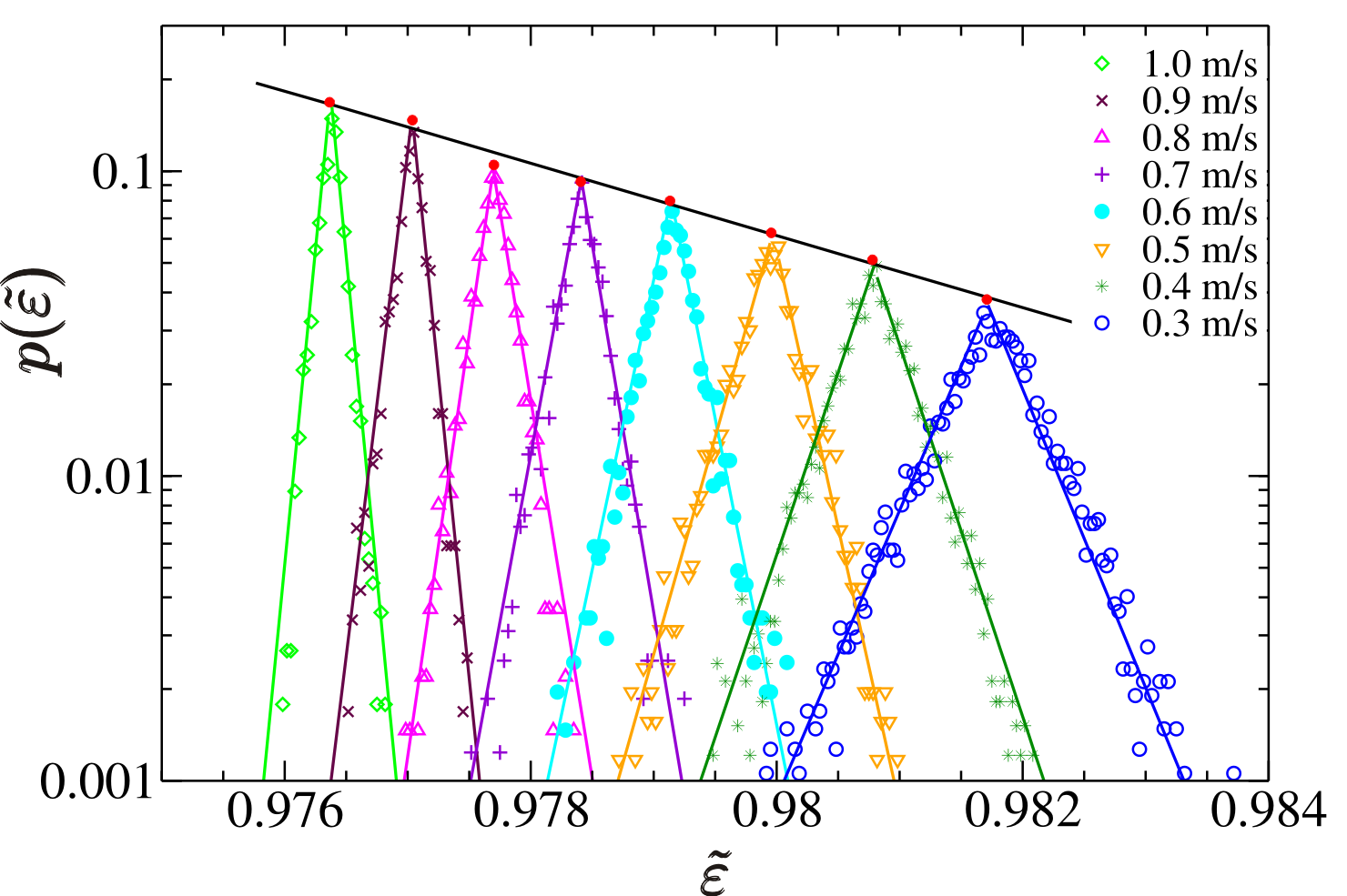}}
\caption{(color online). Simulation results. Top: The {\em macroscopic} coefficient of restitution $\tilde{\varepsilon}$ plotted against the normal center of mass velocity $v_G$ in moment of impact. The data are colored according to the normalized frequency of occurrences. Bottom: Histograms of $\tilde{\varepsilon}$ for velocities from small intervals, centered around $v_G=0.3\:\ldotp\ldotp1.0$ m/s. The lines are exponential fits.}
 \label{fig:Sim}
\end{figure}

\paragraph{Simulation results,}
On the macroscopic level (neglecting the microscopic asperities at the surface of the particle), the simulation results can be analyzed in the same way as the experimental data. We introduce the {\em macroscopic} coefficient of restitution $\tilde{\varepsilon}$ as the ratio of the post-collisional to pre-collisional center of mass velocities in normal direction 
\begin{equation}
 \label{eq:eps_macro}
 \tilde{\varepsilon}=-\dfrac{\vec{v}^{\,\prime}_G\cdot\vec{n}}{\vec{v}_G\cdot\vec{n}}.
\end{equation}
Surprisingly, the macroscopic interpretation of our simulation results depicted in Fig. \ref{fig:Sim}
shows a striking resemblance with the experimental data. While the actual coefficient of normal restitution $\varepsilon$ is a function of the normal impact velocity described by Eq. \eqref{eq:epsvisco}, the macroscopic coefficient of restitution $\tilde{\varepsilon}$ computed via Eq. \eqref{eq:eps_macro} reveals strong fluctuations in agreement with the experiment.

The probability densities, $p(\tilde{\varepsilon})$, (bottom panel of Fig. \ref{fig:Sim}) are also close to the distribution obtained in the experiment. Their shape is excellently approximated by a combination of two exponential functions, one increasing and one decreasing. The peaks of $p(\tilde{\varepsilon})$ are in line with the values of $\varepsilon(v_G)$ obtained from Eq. \eqref{eq:epsvisco} for the corresponding velocities. 

For the case of an absolutely smooth particle (no asperities), the impact velocity $\vec{v}_c$ at the point of contact is the same as $\vec{v}_G$ and $\tilde{\varepsilon}$ is equal to the coefficient of normal restitution $\varepsilon$. Therefore, in this case, there would be no scatter of the measured values of $\tilde{\varepsilon}$. Consequently a scatter of these data must be attributed to the asperities at the surface of the particle.
From the agreement of the experimental data, Fig. \ref{fig:Exp}, and the simulation data, Fig. \ref{fig:Sim}, it became apparent that the tiny microscopic imperfections of the surface of the otherwise macroscopically smooth particle lead to the characteristic fluctuations of the coefficient of normal restitution.
The slight asymmetry in the shape of the probability density in case of the experimental results, i.e., $p(\varepsilon)$ compared to $p(\tilde{\varepsilon})$, may be caused by some additional dissipative forces that, however, have not be taken into account in our model.

Let us consider the role of the sphere's rotational degrees of freedom which may be interpreted as internal degrees of freedom since they do not enter the computation of the coefficient of restitution, neither in the experiment, Eq. \eqref{eq:epsmeas}, nor in the simulation, Eq. \eqref{eq:eps_macro}. Initially, the particle had no spin and only one degree of freedom in its translational motion. However, due to eccentric impacts caused by asperities, the particle gains some rotation and acquires velocity in horizontal direction. The partition of total kinetic energy in rotational and translational degrees of freedom constantly varies from one impact to another. Thus, the kinetic energy of the linear vertical motion (the only component which enters $\varepsilon$) just before a collision is transformed into energy of the rebound vertical velocity, dissipated energy due to the coefficient of restitution, and changes of the horizontal and rotational velocities. The latter two contribution may be positive or negative, leading to a reduced or increased values of the {\em measured} coefficients of restitution, according to Eqs. \eqref{eq:epsmeas} and \eqref{eq:eps_macro}. Therefore, the particle rotation may be considered as a reservoir of internal energy, leading to fluctuations of the measured coefficient of restitution.

The substantial increase of data scattering with the decrease of impact velocity can be attributed to the growing role of the rotational degrees of freedom in the energy partition. In consecutive collisions the translational energy decreases appreciably due to dissipation, but the amount of energy concentrated in rotation varies only slightly. Thus, the proportion of rotational energy to translational energy increases with the number of particle bounces. The kinetic energy transfers from rotational to translational mode and vice-versa result in a more apparent fluctuation of the coefficient of restitution. As a consequence, the scattering of the restitution coefficient increases as the impact velocity decreases.

\paragraph{Conclusion.} 
We have performed an experimental and numerical study of the coefficient of normal restitution as a function of impact velocity. From about $2.2\times 10^5$ experiments of {\em the same} stainless steel sphere bouncing on a massive horizontal plate we determined experimentally the probability distribution of the fluctuations of the coefficient of normal restitution, $\varepsilon$. We found that $\varepsilon$ increases as the impact velocity decreases. For fixed impact velocity, the probability density of the coefficient of restitution $p(\varepsilon)$ is non-Gaussian. It consists of two exponential functions (one increasing, one decreasing) of different slope. We modelled the particle used in the experiment by a mathematical sphere whose surface is covered by a large number of much smaller spheres (asperities) to simulate a certain roughness. The simulations revealed the same properties of the fluctuations. Since the asperities are the only origin of scatter, we conclude that the experimental observed fluctuations of the coefficient of restitution coefficient are due to microscopic surface roughness of the ball, causing energy transfer between the translational and rotational degrees of freedom.

We thank Knut Reinhardt for technical assistance in setting up the robot and Deutsche Forschungsgemeinschaft (DFG) for funding through the Cluster of Excellence Engineering of Advanced Materials.


\end{document}